# Ultrafast cascade charge transfer in multi bandgap colloidal quantum dot solids enables threshold reduction for optical gain and stimulated emission

Nima Taghipour[‡], Mariona Dalmases[‡], Guy Luke Whitworth[‡], Yongjie Wang[‡], and Gerasimos Konstantatos[‡,†,*]

[‡]ICFO, Institut de Ciències Fotòniques, The Barcelona Institute of Science and Technology, Castelldefels, Barcelona, 08860 Spain

[†]ICREA, Institució Catalana de Recerca i Estudis Avançats, Barcelona, 08010 Spain

*E-mail: gerasimos.konstantatos@icfo.eu

**Abstract**

Achieving low-threshold infrared stimulated emission in solution-processed quantum dots is critical to enable real-life application including photonic integrated circuits (PICs), LIDAR application and optical telecommunication. However, realization of low threshold infrared gain is fundamentally challenging due to high degeneracy of the first emissive state (e.g., 8-fold) and fast Auger recombination. In this letter, we demonstrate ultralow-threshold infrared stimulated emission with an onset of 110 $\mu J.cm^{-2}$ employing cascade charge transfer (CT) in Pb-chalcogenide colloidal quantum dot (CQD) solids. In doing so, we investigate this idea in two different architectures including a mixture of multiband gap CQDs and layer-by-layer (LBL) configuration. Using transient absorption spectroscopy, we show ultrafast cascade CT from large band-gap PbS CQD to small band-gap PbS/PbSSe core/shell CQDs in LBL (~ 2 ps) and mixture (~ 9 ps) configuration. These results indicate the feasibility of using cascade CT as an efficient method to reduce optical gain threshold in CQD solid films.

**Keywords:** Optical telecommunications, infrared, Carrier transferring, Colloidal quantum dots, amplified spontaneous emission

Colloidal quantum dots (CQDs) have been shown as a promising solution-processed lasing medium with emission wavelength ranging from the visible[1–10] to the infrared[11–16]. However, lasing action in CQDs suffers from the multiexciton nature of optical amplification, stemming from non-unity degeneracy of the band-edge state [1,7]. This results in high lasing onsets [1,5,7,8] and fast nonradiative multicarrier Auger process, wherein the released energy from the recombination of an electron-hole pair is transferred to a third carrier [17]. These fundamental implications are even more severe in infrared-emitting Pb-chalcogenide CQDs, which feature an 8-fold electron-hole pair degeneracy of the emissive state[12,14,16]. Recent efforts have led to reduced lasing thresholds by virtue of suppression of the trap-assisted Auger decay at supra-nanocrystalline level [15], and suppression of band-edge Auger process by engineering the shape of potential confinement [11]. The combination of the engineered CQDs with supressed Auger decay and electronic doping have further allowed to achieve low-threshold infrared lasing operating at sub-single exciton level [11].

Yet an alternative approach for lowering the stimulated emission threshold is to use cascade charge/energy transfer with the aim to funnel carriers into the gain medium. In this approach, the active medium consists of the donor species serving as the carrier supplier and emitter species acting as the light amplification medium. Previous experiments have demonstrated that using cascade resonance energy transfer (ET) in organic solids can lower the lasing threshold [18–22]. In these studies, the donor species absorbs the incoming pump light, followed by transfer of the photogenerated carriers to the acceptor species acting as the gain active medium [21]. Inspired by the (ET) mechanism in organics that are characterized by large exciton binding energies, we posited that the analogue mechanism in Pb-chalcogenide CQDs, characterized by very small binding energies (i.e. electrons and holes essentially behave as free carriers), would be a cascade charge transfer (CT) mechanism. In our proposed structure, the active medium comprises a binary heterostructure of small CQDs with large bandgap serving as the carrier supplier for the large CQDs performing as the gain medium. Herein, the donor and the emitter CQDs are formed a type-I heterostructure, in which the photogenerated electrons and holes in the donor CQDs are potentially transferred to the emitter CQDs. Assuming that carrier funnelling process is faster than the stimulated emission in the emitter, these carriers would contribute to the light amplification process. Thus, a portion of the required carriers for satisfying the population inversion condition in the emitter (i.e., net gain) is provided from the donor. This allows to achieve the optical gain at lower thresholds compared to only emitter case. Also, earlier numerical calculations showed that using such an approach in CQDs system can reduce the effective gain threshold by 3-4 times [23].

In this work, we investigate the combined effect of ultrafast cascade CT and supressed Auger recombination on CQD optical gain properties using PbS/PbSSe core/alloyed-shell (C/A-S) structures as the effective active media (i.e., emitter) and larger bandgap PbS CQDs as the carrier supplier species (i.e., donor). We studied the effect of CT in two different configurations: 1) a mixture of the donor – emitter CQDs species 2) a layer-by-layer (LBL) architecture made out of the donor and emitter alternatively. In both cases, we observed an ultrafast cascade CT from the donor to the emitter CQDs with a greater rate compared to that of stimulated emission in the effective active gain medium. We also demonstrate a more than 2.5-fold reduction of the ASE threshold achieved by transferring the carriers with a lifetime of ~ 2 ps in LBL case.

Energy-level diagram of the donor and emitter species are shown in Figure 1a, where the photoexcited electron and hole in the donor are nonradiatively transferred to the emitter. The injection of carriers from the donor to the emitter is assumed to be fast and irreversible. The donor species is PbS CQD having a diameter size of ~ 4 nm, and the emitter is PbS/PbSSe C/A-S CQD having a diameter size of ~ 6 nm. We calculated the position of the valence bad maxima (VBM) of the CQDs using ultraviolet photoelectron spectroscopy (UPS). The conduction band minima (CBM) is calculated by subtracting the optical bandgap. Figure 1b shows the photoluminescence (PL) spectra at very low pump fluence for ligand-exchanged solid films (see Methods). As can be seen in LBL heterostructure, the PL of the donor is quenched while the PL intensity of the emitter CQDs is increased due to efficient carrier funnelling. We note that the PL peak position of the emitters does not change in the LBL device, suggesting the absence of the exciton transfer (i.e., energy transfer) in this structure. Also, the thickness of the donor and emitter layers were measured to be 18-20 nm, in which the center-to-center distance between donor and emitter is much longer than Förster radius [24]. Therefore, in our LBL device, the quenching of the PL corresponds to the highly efficient carrier transfer from the donor layer to the emitter. Our finding is consistent with previous reports in which such an efficient CT has been attributed to the increased electronic coupling strength in ligand-exchanged CQD solids [24,25].

To study the amplified spontaneous emission (ASE) of our proposed architecture, we prepared solid films on alumina substrate (see Methods). In doing so, we pumped the samples by the stripe excitation of a femtosecond laser, in which the emission was collected perpendicular to the excitation axis. In Figure 1c, we demonstrate the cartoons of three different cases that we investigated their optical gain performance in this study. Only emitter device is the reference sample in which the gain medium consists of the emitter CQDs, the gain medium in mix device is a binary blend of the donor and emitter CQDs with 50% content, and the LBL device is made out of alternatingly stacked donor and emitter CQDs. All our devices consist of 9 layers as this has been found to be optimal for gain threshold (see Figure S3 and S4); in particular the LBL device consists of 5 layers of the donor and 4 layers of the emitter CQDs. As an exemplary case, we depict the luminescence spectra of LBL device in Figure 1d for increasing pump fluence (see Figure S2). At low pump fluences, the spontaneous emission dominates the PL spectra having a full width at half maxima (FWHM) of ~ 85 nm. While by increasing the pump intensity, an ASE peak appears at a longer wavelength of 1660 nm with a narrower FWHM of ~ 22 nm. The corresponding integrated intensity of ASE peak as a function of pump fluence for different devices are shown in Figure 1e, in which the ASE threshold for the only emitter device is measured as ~315 $\mu J.cm^{-2}$, while the ASE threshold for the mix device is quantified to be ~ 230 $\mu J.cm^{-2}$. Interestingly, the ASE threshold in the LBL device drops to 110 $\mu J.cm^{-2}$. To the best of our knowledge, our obtained ASE threshold in this work is the record lowest threshold among all undoped infrared-emitting CQDs, in which the ASE threshold for the similar sizes PbS core CQDs were reported as 700-800 $\mu J.cm^{-2}$ [14,16] We attribute the lower ASE threshold for the LBL sample over the mix, being due to a higher CT efficiency in the LBL case as donor and acceptor are guaranteed to be adjacent, whereas the mixture most likely forms donor and acceptor rich domains, preventing sufficiently fast charge transfer. [25].

Additionally, we quantified net modal gain coefficient (G) for LBL device using variable stripe length (VSL) technique (see Supplementary Note.1). We depict the results in Figure S4 and S5, in which G is measured to be 1230 $cm^{-1}$ at 350 $\mu J.cm^{-2}$ reaching up to 1900 $cm^{-1}$ at 1100 $\mu J.cm^{-2}$. Furthermore, we conducted waveguide loss measurements to extract the optical loss in our devices (see Supplementary Note.2). Figure S6 plots the results of our measurements, in which the total optical loss (α) is computed to be ~ 34 $cm^{-1}$, while α equals to ~53 and ~103 $cm^{-1}$ for mix and emitter-only devices, respectively. We attribute this substantial reduction of the optical loss in mix and LBL devices to decreasing the re-absorption losses of the stimulated emission in these devices. Note, in both cases the amount of emitter species is decreased by 50% in a given volume. Our observation is consistent with previous reports, where they employed similar mechanism for light amplification in organic thin films [21]. Furthermore, we believe that the lower optical loss in LBL device compared to mix stemming from the thin film uniformity. In mix device, the donor species are random distributed causing some scattering losses. While, the film formation in LBL devices is more uniform, in which the emitter CQDs are sandwiched between two donor layers.

To better understand cascade CT characteristics and its effect on optical gain performance, we carried out ultrafast pump-probe transient absorption (TA) spectroscopy (see Methods). The amplitude of the band-edge bleach signal represents the carrier population which maintains for both electron–hole pairs and excitons [26,27]. Using TA, we monitored the differential absorbance ($\Delta\alpha$) of the devices as a function of time and wavelength for a certain pump fluence. We display time evolution of TA spectra for LBL device in Figure 2a as an exemplary case. The other two cases are provided in Figure S10. For these measurements, we used an excitation wavelength of 750 nm with very low excitation density ($\langle N \rangle$ = 0.001, where $\langle N \rangle$ average number of excitons per dot) in order to minimize any multiexciton recombination but still acquire enough signal-to-noise ratio. We note that the TA spectra shows resonance peaks at 995, 1125 and 1245 nm, which are indicated by the blue arrows in Figure 2a. The peak at 1125 represents the band-edge resonance peak of the donor and higher excited state transition of the emitter CQDs (see Figure S10a&b). Also, the other two bleach peaks represent the higher excited resonances of the emitter (Figure S10b). Especially, the resonance peak at 1125 nm is responsible for the cascade CT process, in which the amplitude of $\Delta \alpha$ is reduced by 16% at early time window 1-10 ps, and it drops to 66% of its initial value after 100 ps. In contrast, we do not observe similar behaviour in donor-only device at the same photoexcitation condition (Figure S10a). Figure 2b plots the kinetic decay of the donor-only device at the exciton resonance peak of 1120 nm, the trace decays with a single time constant of about 5.4 ns. Note, the amplitude of $\Delta \alpha$ in donor-only case is reduced by 16% after 1200 ps, while in LBL case this value is less than 10 ps. As a control sample, we monitored the kinetic traces (see Figure 2c) of the emitter-only at the same wavelength of the donor excitonic resonance. The decay follows a mono-exponential behaviour with a lifetime of 3.2 ns, proving that emitter-only species does not show any fast component at 1120 nm.

We exhibit the kinetic trace of the LBL sample at the resonance peak of 1125 nm in Figure 2d, in which the decay shows a clearly different tendency compared to only-donor and - emitter cases. The $\Delta \alpha$ signal decays with two-time constants having a lifetime of 1.6 and 1900 ps. Figure 2e shows the kinetic decay of the mix device in which the content of the donor and acceptor is 50%. The decay follows bi-exponential function with lifetimes of 9 and 2200 ps. We attribute the appearance of the fast component at early time to the carrier funnelling from the donor species to the emitters. Similar CT have been reported using nanocrystals (NCs) in LBL solid films [28,29]. As we discussed above, the CT should be faster than stimulated emission of the emitters for reducing the optical gain threshold. The results shown in Figure 2 prove that CT in both cases is faster than the stimulated emission of the emitters with a typical lifetime of ~40-50 ps [11,16]. We note that the CT in the LBL is faster than it is in the mix device, causing lower ASE threshold in former case (see Figure 1e). We interpret the slower CT in mix device to the distribution profile of the emitters in the donor matrix, which are randomly distributed throughout the device. This results in the combination of hopping carriers between donors and CT from the donor to the emitters leading to slow CT. However, in the LBL device, the photodegraded charges in the donor species are rapidly hopping in the donor layer and finally are transferred to the emitter layer.

Finally, to shed further insights on the effect of carrier funnelling on the gain threshold, we quantified the optical gain threshold employing TA. In doing so, we excited the LBL device at two different pumping wavelengths, namely 1160 nm and 1500 nm. In former, both donor and emitter species are excited where the optical gain mechanism is CT-assisted concept. In latter, only emitter CQDs are excited in which the optical gain is conventional mechanism. We show the linear absorbance ($\alpha_0$) of the LBL device in Figure 3a, in which the excitation wavelengths are indicated by the arrows. The absorbance of the thin film at 1160 nm and 1500 nm is calculated to be 0.11 and 0.036 cm$^{-1}$, correspondingly. To quantify the gain threshold, we calculate the excited-state absorption $\alpha = \Delta\alpha + \alpha_0$, where $\alpha_0$ is the absorbance of the unexcited sample and $\Delta\alpha$ is the pump-induced absorbance changes at the corresponding ASE wavelength.

Figure 3b plots $\alpha$ as a function of the pumping intensity at the excitation wavelength of 1160 nm, where the blue shading indicates the region of the optical gain ($\alpha < 0$). Then, the gain threshold is calculated to be <87 $\mu$J.cm$^{-2}$ at the excitation wavelength of 1160 nm. This value is consistent with our observed ASE threshold (~110 $\mu$J.cm$^{-2}$) where the pumping wavelength is 1030 nm. We depict $\alpha$ at the excitation wavelength of 1500 nm in Figure 3c, in which the optical gain threshold is quantified as ~520 $\mu$J.cm$^{-2}$. It should be pointed out that the gain threshold in the case of CT-assisted concept is 6 times less than the conventional gain mechanism. While, the absorbed light at 1500 nm is only ~3 times less compared to its value at 1160 nm. Hence, the obtained gain threshold is not only originated from the higher absorption in LBL device, but also the CT from the donor layer to the emitter allows us to reduce the gain threshold.

In this work we demonstrate that charge transfer in CQD films with an engineered potential landscape leads to a drastic reduction in the stimulated emission threshold using. We practically demonstrate ultrafast cascade CT in donor – emitter heterostructure in both mixture and LBL architecture. We also show a considerable, more than 2.5-fold reduction in the ASE threshold owing to ultrafast carrier funnelling from the donor species to the emitter. Moreover, from TA measurements we show that the reduction in the optical gain threshold stems from higher absorption at the excitation wavelength combined with efficient and ultrafast carrier funnelling.

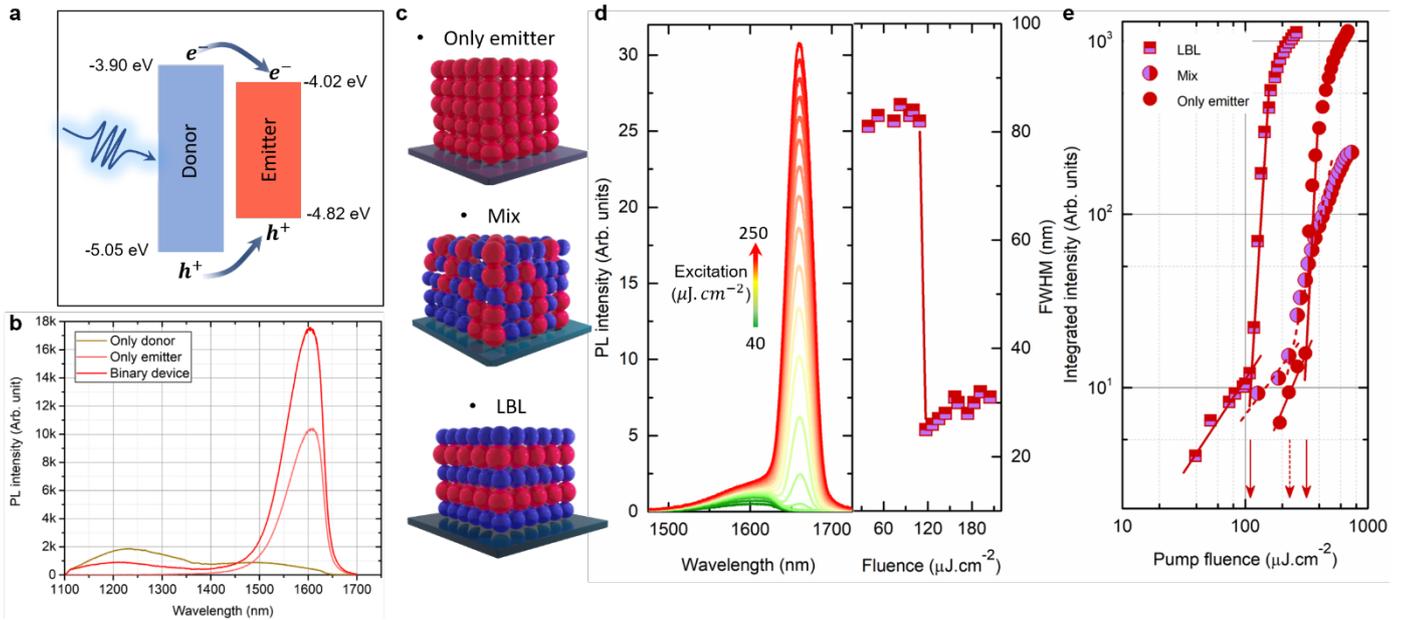

**Figure 1. (a)** Energy-level diagram of the donor- emitter heterostructure. **(b)** PL spectra for different cases where the PL emission from donor CQD quenches and emission intensity of the emitter is enhanced. **(c)** Schematic representation of the investigated devices in this study. The red spheres represent PbS/PbSSe C/A-S CQDs and blue spheres are symbolized large bandgap PbS CQDs. **(d)** ASE spectra of LBL device as a function of pump fluence along with corresponding FWHM. **(e)** Integrated PL spectra depicted as a function of pump fluence at the optical excitation wavelength (1030 nm).

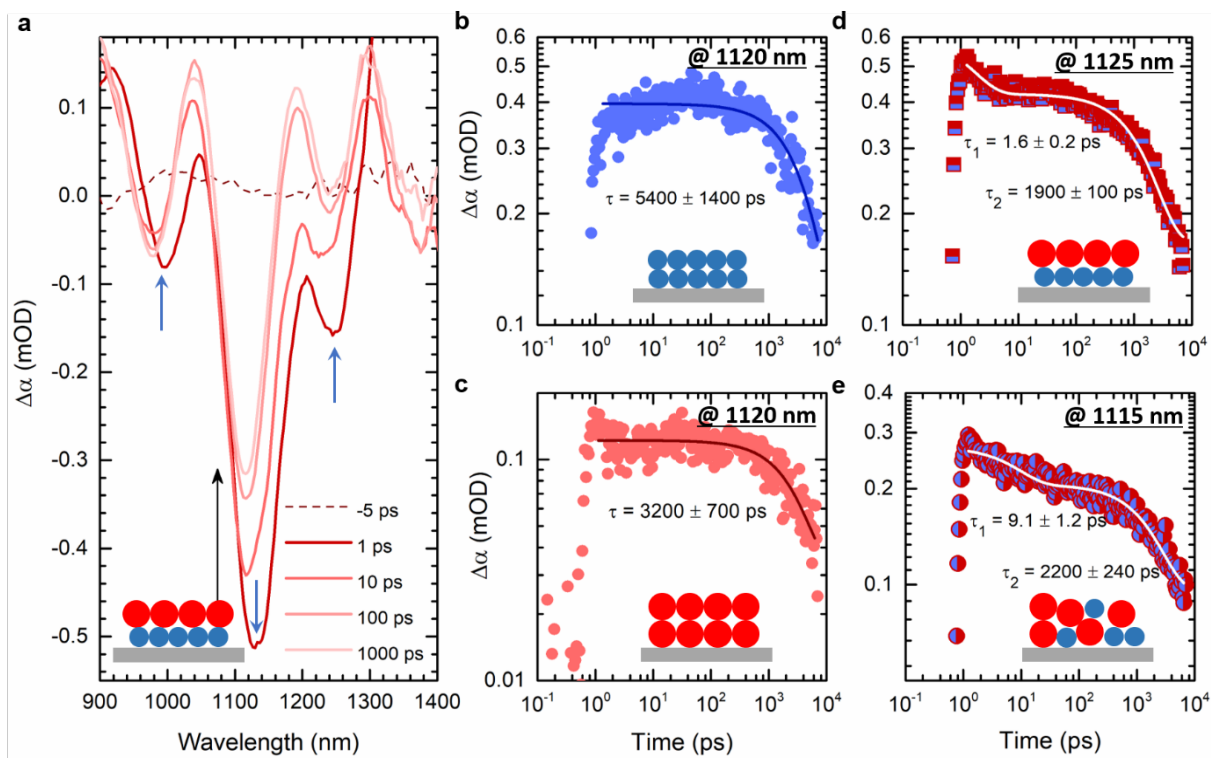

**Figure 2. (a)** Transient absorption spectra at different delay times for LBL device. Time traces at the bleach signal for different cases: **(b)** only donor, **(c)** only emitter, **(d)** LBL and **(e)** mixture device. The solid lines show the exponential fittings. For clarity, the raw data in these traces are smoothed. Fits were obtained from the unsmoothed traces. The inset shows the schematic of corresponding device.

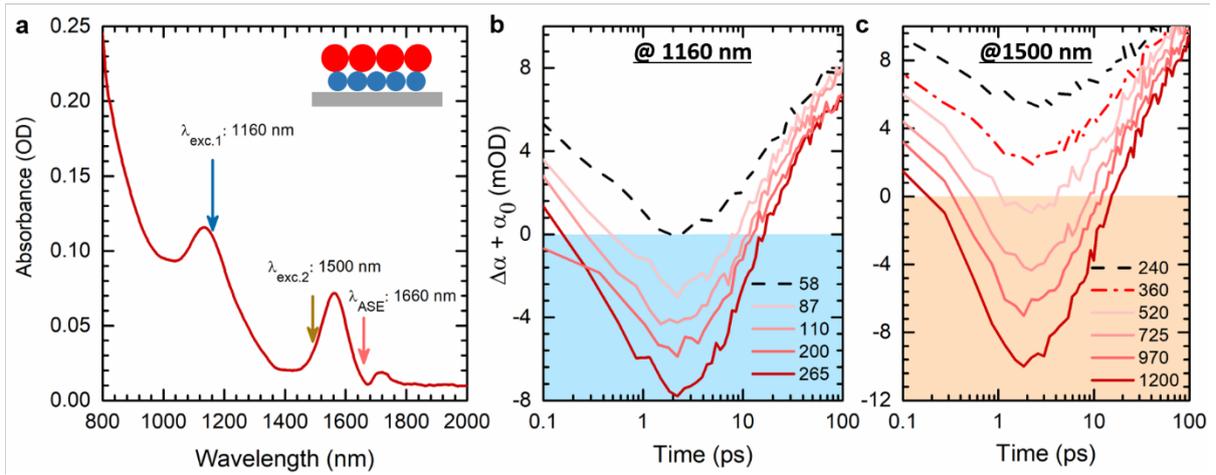

Figure 3. **(a)** Absorbance spectrum for LBL device, in which the excitation wavelengths are indicated by blue- (1160) and green- arrow (1500 nm). The selected wavelength for calculating gain threshold is indicated by red- arrow (1650 nm), which is ASE wavelength. Calculated time-resolved excited-state absorbance $\alpha = \Delta\alpha + \alpha_0$ as a function of pump fluences for LBL device at various excitation wavelength of **(b)** 1160 nm, and **(c)** 1500 nm. Blue and green shading indicate the optical gain region ($\alpha < 0$). Inset show the pump fluence ($\mu J.cm^{-2}$).

## Methods

**Synthesis PbS matrix QDs**: 1100nm excitonic peak PbS QDs were synthesized under inert atmosphere by hot injection method. 446 mg lead (II) oxide (PbO), 18 ml 1-octadecene (ODE) and 6.4 mL oleic acid (OA) were degassed for 1h under vacuum at 100°C. Once under argon, the temperature was set at 95°C and 210µL hexamethyldisilathiane (HMS) in 5mL ODE was quickly injected. After the injection the reaction was cooled down naturally to room temperature. The PbS QDs were purified by dispersion/precipitation with anhydrous toluene/acetone three times. Finally, the concentration was adjusted to 30 mg. mL and the sample was stored at low temperature to avoid Ostwald ripening.

**Synthesis of PbS@PbSe core-shell QDs:** The synthesis of PbS@PbSe core-shell QDs was described previously (ref. C/S paper). Briefly, 446 mg PbO, 50 ml ODE and 3.8 mL OA were heated at 100°C under vacuum for 1h to form lead oleate. Under argon, and keeping the same temperature, a solution of 75µL HMS dissolved in 3 mL ODE was quickly injected. After 6 minutes of reaction a second solution of 115µL HMS in 9 mL ODE was dropwise injected for 12 min. Afterwards, the solution was let cool down naturally. Once at room temperature it was washed three times with a mixture of acetone/ethanol, redispersing in toluene. The final sample was redispersed in degassed ODE and the concentration was adjusted to 70 mg/mL. This solution was stored at low temperature and under nitrogen. To grow a PbSe shell on these PbS core QDs, 240 mg PbO, 10 mL ODE and 1.1 mL OA were heated to 100°C under vacuum, for 1h, to form the lead precursor. After, the solution was changed to argon atmosphere, the temperature was set to 140°C and 2 mL of PbS-ODE solution were injected. When the temperature was recovered, the selenium precursor was injected (96 mg Se powder dissolved in 1.3 mL tributylphosphine). The growth of the shell can be monitored by absorption spectroscopy. When the thickness of PbSe shell was the appropriated, the reaction was quenched with a water bath. The QDs were washed three times with ethanol, redispersing in toluene. Finally, the concentration was adjusted to 30 mg. mL$^{-1}$.

**Thin film deposition.** The active medium was deposited on top pre-cleaned Alumina substrate using spin coating technique. First, a 30 mg/mL of the CQDs in toluene was poured on substrate and wait for 5 s and then spin coating started for 20 s at a speed of 2500 r.p.m. Second, a mixed ligand consists of 7 mg/mL EMII in methanol with 0.035% MPA was poured on the formed solid film and wait for 30 s. Then, the spin coating started for 40 s, where a few drops of methanol poured during the spin coating in order to remove the undesired organic ligands. The mentioned steps were repeated till reaching the desired solid film thickness.

**TA measurements.** The characterizations were conducted using a mode-lock Ti:sapphire femtosecond laser (45 fs) operating at 800 nm and a repetition rate of 1 kHz. To tune the pump wavelength, an optical parametric amplifier used with approximately 3.5 mJ, where the pump fluence is controlled by a neutral density. The optical setup was designed as noncollinear configuration. To control the delay time between pump and probe, a precise motorized stage was used. We employed two different probe signal windows in our TA measurements. For generating probe beam ranging from 800 to 1600 nm, the pump beam excites a nonlinear crystal to generate the white light. To generate the range of 1620 – 2300, another optical parametric amplifier was utilized. In both cases, the probe beam was focused on the sample, in which the pump beam is larger than excitation beam area.

**ASE characterization.** The sample were optically pumped by a femtosecond (300 fs) Yt:YAG ORAGAMI laser operating at a wavelength of 1030 nm with a reputation rate of 10 kHz. For adjusting the pump intensity, we used a variable neutral density. The laser was focused on the sample through a cylindrical lens. The ASE signal was collected perpendicular to the excitation direction using a lens with a focal length of 50 mm having a diameter of 2 inches. The collected ASE spectra was coupled into a Shamrock spectrograph (Andor) which equipped with a 1D-InGaAs camera a lens having 200 mm focal length through 100 μm slit.

**SCAPS simulation.** The schematics of the simulated structures are given in Figure S7. For the illumination condition, we used a monochromatic excitation operates at 1030 nm having a power density of $3 \times 10^4$ W.cm$^{-2}$. The basic parameters for the donor and emitter that we used for the simulation are shown in Table S1.

**Supporting Information**

Absorption spectra, additional ASE spectra, VSL characterizations, waveguide loss coefficient measurements, SCAPS simulation, additional TA spectra.

**Acknowledgements**

G.K acknowledges support from H2020 ERC program grant number 101002306. The authors acknowledge support from the State Research Agency (AEI)/ PID2020-112591RB-I00/10.13039/501100011033 and PDC2021-120733-I00 funded by MCIN/AEI/10.13039/501100011033 by the European Union Next Generation EU/PRTR. Additionally, this project has received funding from the Spanish State Research Agency, through the CEX2019-000910 [MCIN/AEI/10.13039/501100011033], the CERCA Programme/Generalitat de Catalunya and Fundació Mir-Puig. This project has received funding from the European Union's Horizon 2020 research and innovation programme under the Marie Skłodowska-Curie grant agreement number 754558.

**Authors Contribution**

TOC Graphic:

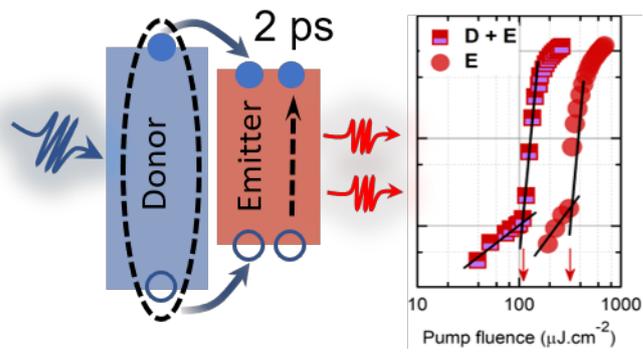